\documentclass[prb,twocolumn,preprintnumbers,amsmath,amssymb,floatfix,longbibliography,superscriptaddress]{revtex4-1}
\usepackage{graphicx}
\usepackage{graphics}
\usepackage{psfrag}
\usepackage{hyperref}
\usepackage{multirow}
\usepackage[sort&compress]{natbib}
\usepackage{amssymb}
\usepackage{amsmath}
\usepackage{amsthm}
\usepackage{bbold}
\usepackage{bbm}
\usepackage{enumerate}
\usepackage{textcomp}

\usepackage{amstext} % for \text macro
\usepackage{array}   % for \newcolumntype macro
\newcolumntype{L}{>{$}l<{$}} % math-mode version of "l" column type

\usepackage{ulem} % to cross out text

\newcommand{\rmi}{{\rm i}}

\DeclareMathOperator{\Tr}{Tr}

\usepackage{xcolor}

\usepackage{hyperref}
\usepackage[figure,table]{hypcap}               % to correct a problem with hyperref
\hypersetup{
pdftitle = {},
pdfsubject = {},
pdfauthor = {},
pdfkeywords = {},
pdfcreator = {},
pdfproducer = {LaTeX with hyperref},
colorlinks = true,
linkcolor = blue,
anchorcolor = blue,
citecolor = blue,
filecolor = red,
menucolor = red,
pagecolor = red,
urlcolor  = blue,
breaklinks = true,
pdfstartview = FitV,
pdfhighlight = /I,
pdfpagelayout = TwoColumn,
hypertexnames=true
}

\begin{document}

\hypersetup{pdftitle={title}}
\title{Time reversal symmetry breaking and $d$-wave superconductivity of triple-point fermions}

\author{Subrata Mandal}
\email{subratam@sfu.ca}
\affiliation{ Department of Physics, Simon Fraser University, Burnaby, British Columbia, Canada V5A 1S6 }

\author{Julia M. Link}
\email{jmlink@sfu.ca}
\affiliation{ Department of Physics, Simon Fraser University, Burnaby, British Columbia, Canada V5A 1S6 }

\author{Igor F. Herbut}
\email{iherbut@sfu.ca}
\affiliation{ Department of Physics, Simon Fraser University, Burnaby, British Columbia, Canada V5A 1S6 }

\begin{abstract}
We study the possibility of complex tensor ($d$-wave) superconducting order in three-dimensional semimetals with chiral spin-1/2 triple-point fermions, which  have an effective orbital angular momentum of $L=1$ arising from a crossing of three bands. Retaining the first three lowest order terms in momentum and assuming rotational symmetry we show that the resulting mean-field $d$-wave ground state breaks time reversal symmetry, but then depends crucially on the coefficients of the two quadratic terms in the Hamiltonian. The phase diagram at a finite chemical potential displays both the ``cyclic" and the ``ferromagnetic" superconducting states, distinguished by the average value of the magnetization; in the former state it is minimal (zero), whereas in the latter it is maximal (two). In both states we find mini Bogoliubov-Fermi surfaces in the quasiparticle spectrum, conforming to recent general arguments. \\[2ex]
\end{abstract}

\maketitle

\section{Introduction}

Crystals' space symmetries allow multiband crossings that lead to topologically non-trivial bands and describe low-energy fermions with effective higher angular momentum \cite{Bradlynaaf5037}. Such exotic fermions provide a fascinating new area of condensed matter physics, and naturally lead to exotic superconductivity, for example, since fermions with higher angular momentum  can obviously be Cooper-paired in various ways. Some
recent examples of semimetals with such higher effective angular momentum that can lead to unconventional superconductivity are the Rarita-Schwinger-Weyl  (RSW) \cite{PhysRevB.101.184503} and the Luttinger semimetals\cite{usc1,usc2,PhysRevB.93.205138,usc4,usc5,usc6,agterberg2017bogoliubov,usc8,usc9,
usc10,PhysRevLett.120.057002,usc12, usc13,usc14,usc15,sim2019topological, usc17,usc18,usc19}, with spin-orbit coupled fermions with the total angular momentum of $L=3/2$.

Both the RSW and Luttinger semimetals have crossing of four energy bands. Three-band crossings\cite{Fulga,Maxwellfermions,Spin-Tensor-Momentum}, however, are also possible, and were recently observed in CoSi\cite{Rao2019ObservationOU} and signs of superconductivity in such materials were observed in ${\mathrm{PdSb}}_{2}$ \cite{PhysRevB.99.161110}. The quasiparticles participating in the triple band crossing appear as having an effective orbital angular momentum of $L=1$ and, as we will discuss, can therefore form local Cooper pairs with a total angular momentum of $J=0,1,2$, i. e. exhibit $s$-, $p$-, and $d$-wave superconductivity, respectively. The $J=0$ channel was already studied by Lin and Nandkishore in ref.~\onlinecite{PhysRevB.97.134521} and $J=1$ vector pairing of spinless fermions was examined in ref.~ \onlinecite{PhysRevResearch.2.043209,sim2019tripletsuperconductivity}. The three-component $p$-wave ($J=1$) and the five-component $d$-wave ($J=2$) superconducting order parameters  are particularly interesting, since they offer a possibility of the superconducting state breaking the time reversal (TR) symmetry, and thus manifesting some form of magnetism in coexistence with superconductivity. The case of tensorial $d$-wave pairing is in this context especially rich. The possibility of $d$-wave superconducting order of triple-point fermions has to our knowledge not yet been studied, although it has enjoyed a long history in relation to the physics of neutron stars\cite{SaulsSerene} and $^3 \text{He}$ \cite{mermin1974d}, and more recently, of Bose-Einstein condensates \cite{KawaguchiUeda} and Luttinger semimetals \cite{usc1,usc2,PhysRevB.93.205138,usc4,usc5,usc6,agterberg2017bogoliubov,usc8,usc9, usc10,PhysRevLett.120.057002,usc12, usc13,usc14,usc15,sim2019topological, usc17,usc18,usc19}. Closing this gap in the growing literature on the subject and at the same time continuing our systematic study of the $d$-wave superconducting order in various physical settings, we here focus entirely on the $J=2$ Cooper pairing of spin-1/2 fermions near a single three-band ($L=1$) crossing. We find an important new feature emerging; in all previous studies the coefficient of one of the three quartic terms in the Ginzburg-Landau theory for the $d$-wave order parameter that discriminated between different TR symmetry-broken superconducting states was either precisely zero, or parametrically small and positive, the latter leading to the ``cyclic" ground state with maximal TR symmetry breaking but zero magnetization. In contrast, in the present case of triple-point fermions we find this crucial Ginzburg-Landau coefficient for the first time to depend nontrivially on the values of the coefficients of the two subleading, rotationally invariant terms in the single-particle Hamiltonian, which are quadratic in momentum. As a result, it can be of either sign. The two superconducting ground states that result from this dependence both break TR symmetry maximally, but differ crucially in their magnetization properties. Whereas the already mentioned cyclic state has the minimal (zero) average magnetization and only shows the quadrupolar magnetic order, the ``ferromagnetic" state shows maximal average magnetization of two. The spin-1/2 triple fermions appear therefore to be the first system that may exhibit the ferromagnetic $d$-wave superconducting state, provided of course that the pairing in the $d$-wave channel dominates over other possibilities.

The paper is organized in the following way. In Sec.~\ref{sec:model} we introduce the Hamiltonian that describes low-energy fermions with effective orbital angular momentum of $L=1$ and spin $S=1/2$. These fermions can form Cooper pairs with total angular momentum $J=0,1,2$, as discussed in Sec.~\ref{sec:complexorderparameter}. The Ginzburg-Landau free energy for the $d$-wave superconducting order parameter is studied at the mean-field level in Sec.~\ref{sec:Ginzburg-Landau}, where the potential ground states are introduced prior to the calculation of the coefficients of the free energy, presented in detail in App.~\ref{sec:computation}. In Sec.~\ref{sec:subgroundstate} we discuss how the curvature of the energy band strongly influences the $d$-wave superconducting ground state in the weak coupling regime, and demonstrate that the curvature may be taken as the knob that tunes between the cyclic and the ferromagnetic state. Finally, in Sec.~\ref{sec:discussion} we summarize our findings.

\section{Hamiltonian}
\label{sec:model}
\begin{figure*}[t]
\centering

\includegraphics[width=2\columnwidth]{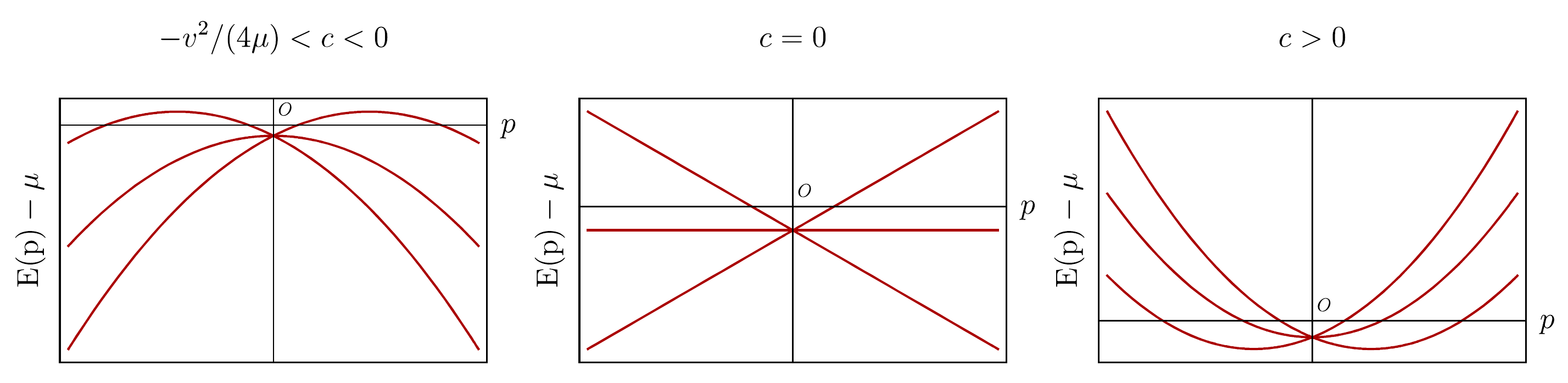}

\caption{The energy dispersion of $H(\textbf{p})$ in the normal state for different values of $c$ when $b=0$ and $\mu>0$. Depending on the curvature of the energy bands the number of Fermi surfaces varies from two for $-v^2/(4\mu)<c<0$, to one for $c=0$, and three for $c>0$.}
\label{fig:Fig1}
\end{figure*}

We consider the system of spin-1/2 fermions  with the low-energy spectrum of the lattice Hamiltonian exhibiting a crossing of three bands \cite{Bradlynaaf5037} at the center of the Brillouin zone:
\begin{equation}
 H(\boldsymbol{p})=
 \mathbb{1}_{2\times 2} \otimes (H_{0}(\boldsymbol{p}) -\mu \: \mathbb{1}_{3\times 3})
 \:.
 \label{eq:Eq1}
\end{equation}
We can think of the left factor in the tensor product as acting on the spin-like degree of freedom, and the $H_0$ as acting on the orbital-like degree of freedom. $\mu$ is the usual chemical potential. For a crystal with cubic or tetragonal symmetry the dynamics of the fermions near the crossing is described by the following TR-symmetric Hamiltonian, expanded to the second order in momentum:
\begin{equation}
H_0(\boldsymbol{p})
=
%[
v\:\boldsymbol{p}\cdot \boldsymbol{L} + c\: p^2 \mathbb{1}_{3 \times 3}  + b\: (\boldsymbol{p}\cdot \boldsymbol{L})^2
% ]
 \:.
\end{equation}
The Hamiltonian effectively describes particles with angular momentum of $L=1$, with
the three dimensional matrices $L_i$, $i=1,2,3=x,y,z$, that form the spin-1 representation of the Lie algebra of the $SO(3)$ group of rotations, with $[L_i,L_j]=\rmi \epsilon_{ijk}L_k$. In the standard representation these are given by the matrices
\begin{align}
 L_x &=\frac{1}{\sqrt{2}}
\begin{pmatrix}
 0 & 1 & 0 \\
 1 & 0 & 1 \\
 0 & 1 & 0 \\
\end{pmatrix} \:, \\
L_y &= \frac{\rmi}{\sqrt{2}}
\begin{pmatrix}
 0 & -1 & 0 \\
 1 & 0 & -1 \\
 0 & 1 & 0 \\
\end{pmatrix} \:,\\
L_z &= \begin{pmatrix}
 1 & 0 & 0 \\
 0 & 0 & 0 \\
 0 & 0 & -1 \\
\end{pmatrix}\:.
\end{align}
The low-energy Hamiltonian $H_0$ is the most general such Hamiltonian to the second order in momentum, measured from the crossing. It also has a full rotational symmetry, which we are assuming here for simplicity. The first (linear) term of the Hamiltonian, proportional to the velocity $v>0$, breaks the inversion symmetry, and exhibits the crossing of two energy bands linear in momentum and of a flat band (see Fig.~\ref{fig:Fig1}). The two remaining distinct second-order terms add curvature to all three energy bands. The energy dispersion of the bands is given by:
\begin{eqnarray}
 E_{\pm 1}&=&(b+ c) p^2 \pm v |p| -\mu \:,\\
 E_0&=&c p^2 -\mu
 \:.
\end{eqnarray}
Depending on the curvatures of the bands, set by the parameters $b$ and $c$, the number of Fermi surfaces at a chemical potential $\mu >0$ varies from zero to three, as can be seen in Fig.~\ref{fig:Fig1}, where as an illustration we set $b=0$ and varied only $c$. When $b+c<-\frac{v^2}{4\mu}$ and $c<0$ no energy band crosses the chemical potential and no normal Fermi surface emerges, whereas for $-\frac{v^2}{4\mu}<b+c$ and $c<0$ the energy bands cross the Fermi level twice, which leads to two Fermi surfaces. At the special point $b=0,c=0$ only one Fermi surface appears, and finally for $b+c>-\frac{v^2}{4\mu}$ with $c>0$ we find all three Fermi surfaces. The Fermi momenta and the Fermi velocities are generally given by

\begin{eqnarray}
 & k_{{\rm F},+1_\pm}  = \frac{-v \pm \sqrt{v^2+4(b+c)\mu}}{2(b+c)}\:, \quad &v_{{\rm F},+1_\pm } = \sqrt{v^2 +4 (b+c) \mu} \label{eq:kf1} \\
 & k_{{\rm F},0} = \sqrt{\frac{\mu}{c}} \:, \quad &v_{{\rm F},0} = 2 \sqrt{c \mu} \label{eq:kf0}\\
 & k_{{\rm F},-1_\pm} =  \frac{v\pm\sqrt{v^2+4(b+c)\mu}}{2(b+c)}\:, \quad &v_{{\rm F},-1_\pm} = \sqrt{v^2 +4 (b+c) \mu} \label{eq:kfm1}
 \:.
\end{eqnarray}

An exception occurs when $b=-c$: $E_{\pm 1}$ become then linearly dispersive, but $E_0$  remains parabolic.  As a result, there is always a Fermi surface at $p=|\frac{\mu}{v}|$, and another Fermi surface emerges if $k_{F,0}$ becomes real. In Sec.\ref{sec:subgroundstate} we show that the superconducting ground state of the system crucially depends on the curvature of the energy bands. \\

The Hamiltonian $H_0$ is invariant under TR symmetry and thus commutes with the antiunitary operator $\mathcal{T}_0=U_0 \mathcal{K}$ with $\mathcal{T}_0^2=+1$, where $\mathcal{K}$ is the complex conjugation, and $U_0$ is the unitary matrix
\begin{equation}
 U_0= e^{-\rmi \pi L_y}=\begin{pmatrix}
              0 & 0 & 1\\
              0 & -1 & 0 \\
              1 & 0 & 0
             \end{pmatrix}
\:.
\end{equation}
The TR operator for the full Hamiltonian $H(\boldsymbol{p})$ is given by $\mathcal{T}=\sigma_2 \otimes \mathcal{T}_0=\mathcal{U}\mathcal{K}$ with $\mathcal{U}=\sigma_2\otimes U_0$, and therefore $\mathcal{T}^2=-1$, as appropriate to particles with half-integer spin.

\section{Superconducting order parameter}
\label{sec:complexorderparameter}
\begin{table}
\begin{tabular}{c c c c}
\hline
\hline
$S$ & $L$ & $J$ & $M_J$ \\
\hline
$0$ & $0$ & $0$ & $\mathbb{1}_{2\times 2} \otimes \mathbb{1}_{3\times 3}$\\
$1$ & $1$ & $0,1,2$& $\sigma_i \otimes L_j$ \\
$0$ & $2$ & $2$ & $\Gamma_a =\mathbb{1}_{2\times2} \otimes \gamma_a $\\
\hline
\hline
\end{tabular}
\caption{The allowed local superconducting pairings with the Cooper pair's total angular momentum $J$, total spin $S$ and orbital angular momentum $L$, and their corresponding pairing matrices $M_J$. $\sigma_i$ are the Pauli matrices and $\gamma_a$ are defined as $\gamma_1= L_x^2-L_y^2$, $\gamma_2=\frac{1}{\sqrt{3}}(2 L_z^2-L_x^2-L_y^2)$, $\gamma_3=L_x L_z+L_z L_x$, $\gamma_4= L_y L_z+L_z L_y$, $\gamma_5= L_x L_y+L_y L_x$.}
\label{tab:Tab1}
\end{table}
We consider next the local channels available for Cooper pairing of the triple-point fermions.
The chiral fermions with orbital angular momentum of $L=1$ and spin $1/2$ can form Cooper pairs with total-angular momentum $J=0,1,2$, i.e. the pairing channels consist of $s$-, $p$-, or $d$-wave order parameter. More precisely, for two fermions with $\big(S=1/2 \big)\otimes \big(L=1\big)$ we have the standard angular momentum algebra
\begin{equation}
 \big( \frac{1}{2} \otimes 1 \big) \otimes \big( \frac{1}{2} \otimes 1 \big)=
 \big(0\oplus 1 \big) \otimes \big( 0 \oplus 1 \oplus 2 \big)
 \:,
\end{equation}
where the first bracket on the right hand side refers to total spin, and the second to the total orbital angular momentum. Since the electrons obey Fermi statistics the only allowed combinations of total spin and total orbital angular momentum are those that are completely antisymmetric under exchange of particles, and these are $(S,L)$=\{$(0,0)$, $(0,2)$, and $(1,1)$\}, i.e. $s$-, $d$-, and $p$-wave order parameters. If we conveniently consider pairing between time-reversed states \cite{PhysRevB.101.184503}, the pairing matrices $M_J$ that correspond to these allowed channels are then even under time-reversal, i.e. $[M_J,\mathcal{T}]=0$. The different pairing channels and the corresponding matrices that are allowed by the Fermi statistic are listed in table~\ref{tab:Tab1}, where the channel $(S,L,J)=(0,0,0)$ corresponds to the $s$-wave order parameter, $(1,1,J)$ to the $p$-wave order parameter, and $(0,2,2)$ to the $d$-wave order parameter.

In this paper we focus on the $(0,2,2)$ channel, i. e. the spin-singlet $d$-wave order parameter, and neglect possible Cooper pairing in other channels. The simplest Lagrangian yielding the $d$-wave pairing may be written as
\begin{equation}
 L=\psi^\dagger \big( \partial_\tau + H(\textbf{p}) \big) \psi
 - g \big( \psi^\dagger \Gamma_a \mathcal{U}  \psi^* \big) \big( \psi^{\rm T} \mathcal{U} \Gamma_a \psi \big)
 \:,
 \label{eq:Lagrangian1}
\end{equation}
where $\psi(\boldsymbol{x},\tau)=(a_{1,\uparrow},a_{0,\uparrow},
a_{-1,\uparrow},a_{1,\downarrow},a_{0,\downarrow},
a_{-1,\downarrow})^{\rm T}$ is a 6-component Grassmann field, $\tau$ denotes the imaginary time, $\textbf{p}=-i\nabla$ is the momentum operator, the coupling $g>0$, and $\Gamma_a=\mathbb{1}_{2\times2} \otimes \gamma_a$. The sum over repeated indices is assumed.  The matrices $\gamma_a$, $a=1,2,...5$,  are defined  in the caption of Table~\ref{tab:Tab1}. We ignore the issue of a possible physical origin of the pairing interaction and take the coupling $g$ as an effective parameter that leads to a $d$-wave superconducting state. Instead of the dynamical pairing mechanism our problem is the actual nature of the $d$-wave state in the system given the simplest phenomenological interaction that manifestly favors this particular order parameter, but does not distinguish between its different components.

Just as the condensation of the single complex $\Delta_s=\left<\psi^{\rm T}\mathcal{U} \psi \right>$ would correspond to the onset of the conventional $s$-wave superconducting order parameter, the condensation of any linear combination of the five complex $\Delta_a=\left< \psi^{\rm T} \mathcal{U} \Gamma_a \psi \right>$ indicates the onset of the $d$-wave. The explicit expressions for both $\Delta_s$ and $\Delta_a$ in terms of fermion operators are given in Appendix~\ref{ap:cooperpairing}.

The five complex components of $\Delta =\big(\Delta_1, \ldots, \Delta_5 \big)$ and the pairing matrices $\Gamma_a$ transform as $j=2$ irreducible  representation of the $SO(3)$. (See Appendix~\ref{ap:APPB}.) We may therefore arrange the five $\Delta_a$ into a matrix $\phi$, which is an irreducible second-rank tensor under rotations, defined as
\begin{align}
 \label{eq5}  \phi_{ij} = \Delta_a M_{a, ij}.
\end{align}
The five real Gell-Mann matrices \cite{PhysRevB.92.045117} $M_a$ provide a basis of three-dimensional symmetric real traceless matrices. We choose the particular representation in which
\begin{align}
 \label{eq5b}  \phi =\begin{pmatrix} \Delta_1-\frac{1}{\sqrt{3}}\Delta_2 & \Delta_5 & \Delta_3 \\ \Delta_5 & -\Delta_1-\frac{1}{\sqrt{3}} \Delta_2 & \Delta_4 \\ \Delta_3 & \Delta_4 & \frac{2}{\sqrt{3}}\Delta_2 \end{pmatrix}.
\end{align}

\section{Ginzburg-Landau theory}
\label{sec:Ginzburg-Landau}
In the first part of this section we describe the general Ginzburg-Landau free energy for rotationally invariant systems that describes the phase transition to the $d$-wave order parameter, and analyze the possible superconducting ground state configurations that would minimize it. We then discuss the dependence of the actual superconducting ground state in the present case on the values of the parameters $b$ and $c$ which add curvature to the energy bands.

\subsection{Possible superconducting ground states}
\label{sec:Possible superconducting ground states}
The Ginzburg-Landau free energy describing a finite-temperature second-order phase transition towards a $d$-wave order parameter is given by \cite{mermin1974d, PhysRevLett.120.057002}
\begin{align}
\label{mod6} F(\Delta) = F_2(\Delta)+ F_4(\Delta) + \mathcal{O}(\Delta^6),
\end{align}
with the terms quadratic and quartic in the uniform order parameter as
\begin{align}
 \label{mod7} F_2(\Delta) &= r \Delta^* _a \Delta_a,\\
 \label{mod8} F_4(\Delta) &=
 q_1 (\Delta^* _a \Delta_a)^2 +q_2 |\Delta_a \Delta_a|^2 + \frac{q_3}{2} \Tr \big(\phi^\dagger \phi \phi^\dagger \phi \big).
\end{align}
The superconducting phase ensues when the quadratic coefficient $r<0$; the coefficient $q_1$ also needs to be positive for $F$ to be bounded from below. Note that the quartic term multiplied by the coefficient $q_1$ has the same value for all normalized states, so the value of the coefficient $q_1$ does not play a role in the selection of the superconducting order parameter. The signs and magnitudes of the two remaining coefficients $q_2$ and $q_3$ generally decide the broken symmetry state of the system. Let us first understand the role of the coefficient $q_2$, and assume $q_3$ to be sufficiently small.
The value of the product $|\Delta_a \Delta_a|^2 $ in our representation quantifies the overlap of the macroscopic superconducting state with its time-reversed counterpart \cite{PhysRevB.101.184503}. Hence, if $q_2<0$, the real order parameters which describe TR-preserving states maximize this term, and are therefore favored. The matrix $\phi$ can then be rotated into the diagonal form:
\begin{equation}
\label{eqAPPBa}
\phi_{\rm real}= {\Delta}_1 M_1 + {\Delta}_2 M_2
\:,
\end{equation}
 with the relative values of $\Delta_1$ and $\Delta_2$ that need to be determined from higher order terms, or by considering Gaussian fluctuations around the mean-field solution. If, on the other hand, $q_2>0$, the complex states with $ \Delta_a \Delta_a =0$ and with maximally broken TR symmetry are preferred. There are also infinitely many order parameters which break TR symmetry maximally, so $q_2$ alone does not uniquely select the superconducting ground state. This brings us to the role of the remaining coefficient $q_3$. The sign of $q_3$ decides which of the complex superconducting states that break TR maximally is the superconducting ground state. The third term in the right-hand side of Eq.(\ref{mod8}) with the coefficient $q_3$ may be shown to be related to the average magnetization of the state,\cite{PhysRevB.101.184503} so that when $q_2>0$ and $q_3<0$, the state with maximal average magnetization and maximally broken TR symmetry is favored. This is the ``ferromagnetic" state with
\begin{equation}
\label{eqAPPBb}
 \phi_{\rm ferro}=\frac{\Delta}{\sqrt{2}} (M_1 + \rmi M_5 )
 \:.
\end{equation}
However, if the coefficients $q_2$ and $q_3$ both are positive, then the magnetization of the ground states needs to be minimized while keeping the TR symmetry maximally broken. These two requirements are not mutually exclusive, and are in fact fulfilled by the ``cyclic" state, which is defined as
\begin{equation}
\label{eqAPPBc}
 \phi_{\rm cyclic}=\frac{\Delta}{\sqrt{2}}(M_1 + \rmi M_2)
 \:.
\end{equation}
The details of arriving at Eqs.~\eqref{eqAPPBa}, \eqref{eqAPPBb} and
\eqref{eqAPPBc} are presented in Appendix~\eqref{ap:APPB}. The potential ground states configurations can also be systematically identified from a symmetry-based classification\cite{KawaguchiUeda,herbut2019ground} of the $d$-wave order parameter. These states always show some residual symmetry, breaking the rotational $SO(3)$ invariance of the normal state. For instance, the ferromagnetic state and the cyclic state, which will be of our main interest eventually, have the remaining $SO(2)$ and tetragonal symmetry, respectively.
\\

The complete phase diagram for rotationally invariant $d$-wave superconductors was first obtained by Mermin\cite{mermin1974d} and has also been rederived and discussed at length in ref.~\onlinecite{PhysRevB.101.184503}. The one-loop values of the quartic coefficients have been computed in several examples so far. Most importantly, for the spherically symmetric Luttinger semimetals and standard BCS $d$-wave superconductors, $q_3$ has been shown to be exactly zero\cite{mermin1974d,PhysRevLett.120.057002,sim2019topological,herbut2019ground}, whereas for the RSW semimetals $q_3$ turns out to be small and positive \cite{PhysRevB.101.184503}. In the case of three-band crossing under consideration here we show that a qualitatively novel situation arises with $q_3$ having either sign, depending on curvatures of the energy bands.  \\

In the next section we describe the results of the computation of the quartic coefficients for triple-point fermions, and use it to determine the $d$-wave superconducting ground state of the system.

\subsection{Superconducting ground state}
\label{sec:subgroundstate}
\begin{figure}
 \includegraphics[width=\columnwidth]{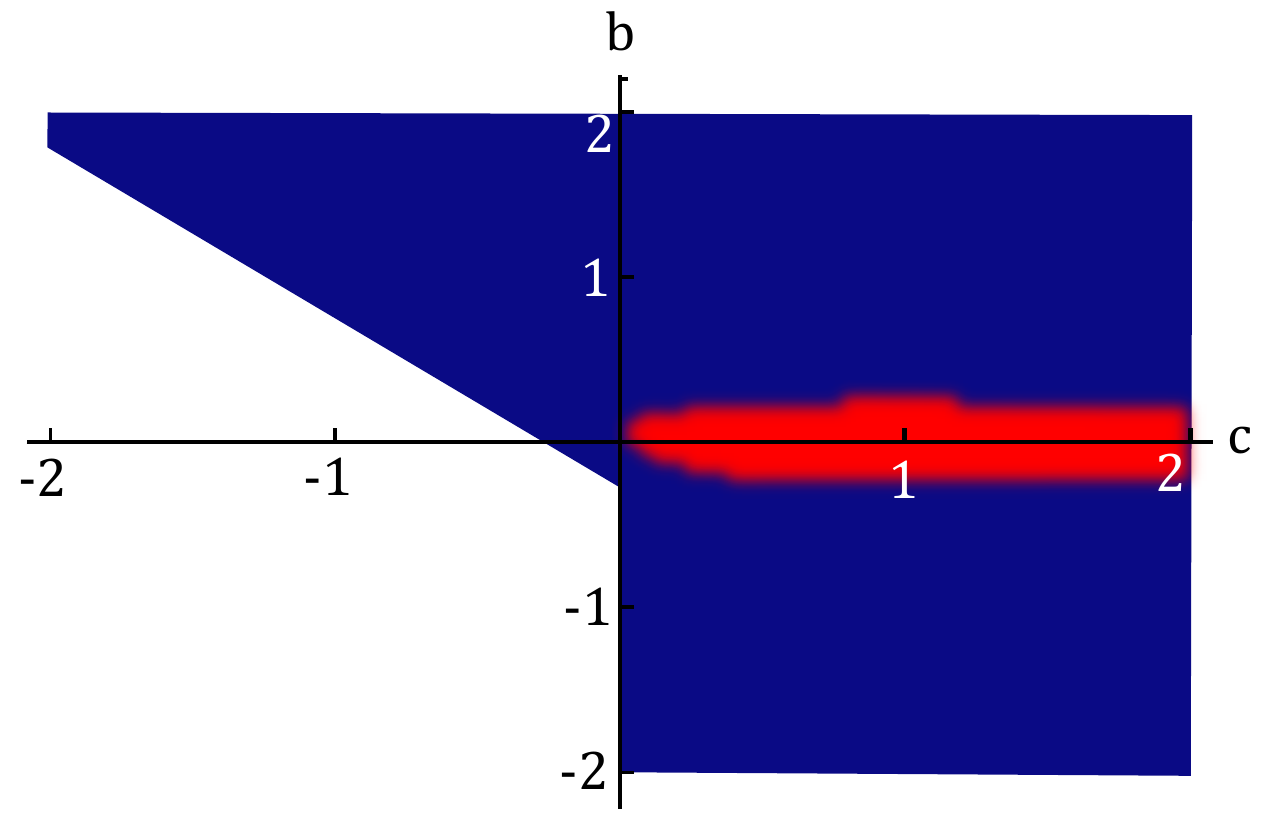}
 \caption{The phase diagram of $d$-wave superconductor in semimetals described by $H(\textbf{p})$, as a function of $c$ and $b$. This plot is obtained for $\mu=1$ and $v=1$. We find two different superconducting ground states that break the time-reversal symmetry maximally: the ferromagnetic state (red) and the cyclic state (blue). The white area corresponds to $-v^2/(4\mu)>c+b$ and $c<0$, when the energy bands do not cross the chemical potential in the normal state.
 }
 \label{fig:Fig3}
\end{figure}
After performing the standard but lengthy one-loop integration over triple-point fermions (details are given in App.~\ref{sec:computation}), we find that for a finite and positive chemical potential the coefficients $q_1$ and $q_2$ are always positive irrespective of the values of the parameters $b$ and $c$ in the  Hamiltonian, and have the standard temperature dependence of $\sim 1/T^{2}$ \cite{StintzingZwerger,igor2000}. In contrast, the coefficient $q_3$ changes its sign depending on the values of $b$ and $c$, as shown in Fig.~\ref{fig:Fig3}, and it depends only logarithmically on temperature.

To understand the different temperature dependence of the quartic terms' coefficients, and in particular why the sign of the coefficient $q_3$ is influenced by $b$ and $c$, let us consider $\mu>0$ and set $b=0$ for simplicity so that only the parameter $c$ determines the curvature of the energy bands. In this case, we find that $q_1$ and $q_2$ are positive for all physically relevant values $-v^2/(4\mu)<c$, whereas the coefficient $q_3$ is only positive if $c \leq 0$ and it is negative for $c>0$. The reason for this change in sign of $q_3$ lies in the analytical structure of the integrands of $q_{1,2,3}$, and, in particular, in the way that this structure depends on the curvature of the energy bands. To this end, we carry out the finite-temperature Matsubara sum of the one-loop integral over fermions that define the coefficients and Taylor expand around the Fermi momenta of the normal state. The coefficients $q_{1,2,3}$ then reduce to the following form:
\begin{widetext}
\begin{equation}\label{q12int}
 q_{1,2}(T,c,b)
 \approx \int_0^\Lambda d k \sum_{i} \bigg( \frac{k_{{\rm F},i}^2}{a^{(1,2)}_{k_{\rm F},i} v_{{\rm F},i}^3 |\boldsymbol{k}-\boldsymbol{k}_{{\rm F},i}|^3 + a^{(1,2)}_{t,i} T^3}
 + \frac{1}{v^2}\frac{1}{b^{(1,2)}_{k_{\rm F},i} v_{{\rm F},i} |\boldsymbol{k}-\boldsymbol{k}_{{\rm F},i}| +b^{(1,2)}_{t,i} T} + \mathcal{O}(|\textbf{k}-\textbf{k}_{\rm F}|^0) \bigg) \:,
\end{equation}
\end{widetext}
and
\begin{widetext}
\begin{equation}\label{q3int}
 q_{3}(T,c,b) \approx \sum_{i} \int_0^\Lambda d k \frac{1}{v^2}\frac{1}{{\tilde{b}}_{k_{\rm F},i} v_{{\rm F},i} |\boldsymbol{k}-\boldsymbol{k}_{{\rm F},i}| +{\tilde{b}}_{t,i} T}
 + \mathcal{O}(|\textbf{k}-\textbf{k}_{\rm F}|^0)
 \:,
\end{equation}
\end{widetext}
where $ \{ a^{(1)}_{k_{\rm F},i}, a^{(1)}_{t,i}, b^{(1)}_{k_{\rm F},i}, b^{(1)}_{t,i} \}$, $ \{ a^{(2)}_{k_{\rm F},i}, a^{(2)}_{t,i}, b^{(2)}_{k_{\rm F},i}, b^{(2)}_{t,i} \}$, and $ \{  {\tilde{b}}_{k_{\rm F},i}, {\tilde{b}}_{t,i} \}$  are  numerical constants associated with quartic coefficients $q_1$, $q_2$, and $q_3$, respectively. Here, the sum is over all Fermi surfaces indexed by $i$, and $k_{{\rm F},i}$ and $v_{{\rm F},i}$ represent the Fermi momentum and Fermi velocity of the corresponding Fermi surface.

Finally, performing the momentum integration in Eq.~\eqref{q12int}, we find that $q_{1,2}$ acquire the usual $\sim 1/T^{2}$ temperature dependence to the leading order in inverse temperature, with the next order correction proportional to $\log(T)$. In the coefficient $q_3$, however, the leading order term in the integrand that would be of the form of $({\tilde{a}}_{k_{\rm F},i} v^3_{{\rm F},i} {|\textbf{k}-\textbf{k}_{{\rm F},i}|}^3+{\tilde{a}}_{t,i}T^3)^{-1}$ is absent, leaving the first non-zero term in the Taylor expansion proportional to $({\tilde{b}}_{k_{\rm F},i} v_{{{\rm F},i}} |\textbf{k}-\textbf{k}_{{\rm F},i}|+{\tilde{b}}_{t,i}T)^{-1}$. The absence of the leading order term causes the different temperature dependence of $q_{1,2}$ and $q_{3}$, and it was also found in the example of $d$-wave pairing of the RSW fermions in ref.~\onlinecite{PhysRevB.101.184503}. We find that the coefficients near the critical temperature ($T_c$) are given by the following expressions to the leading order in $\mu/T_c \ll 1$:
\begin{equation}\label{q1q2form}
 q_{1,2}(T_c,c,b) = \sum_{i} {\eta}^{(1,2)}_{i} \frac{k_{{\rm F},i}^2}{v_{{\rm F},i}} \frac{1}{T_c ^ 2} + \mathcal{O}(\log(\mu/T_c))
\end{equation}
and
\begin{equation}\label{q3form}
 q_3(T_c,c,b) =  \frac{1}{v^2}\sum_{i} \tilde{\eta}_{i}\frac{1}{v_{{\rm F},i}} \log(\mu/T_c )
 \:,
\end{equation}
where ${\eta}^{(1)}_i$, ${\eta}^{(2)}_i$, and $\tilde{\eta}_i$ are numerical coefficients.

What still remains to be explained is the dependence of the coefficient $q_3$ on parameters $b$ and $c$. As we already noted, parameters $b$ and $c$ determine which energy bands cross the Fermi level and thus control the number of Fermi surfaces in the normal state. Hence, the parameter $c$ sets the number of terms that contribute to the coefficients when we set $b=0$. For example, if $c < 0$, there are two Fermi surfaces, and to obtain the quartic coefficients one needs to sum over $i=+1_+ , +1_-$. In other words, the coefficients then receive two contributions with the same Fermi velocity $v_{{\rm F},+1_+}=v_{{\rm F},+1_-}$ but with different Fermi momenta $k_{{\rm F},+1_+}$ and $k_{{\rm F},+1_-}$, as defined in Eqs.~\eqref{eq:kf1}-\eqref{eq:kfm1}. If $c=0$, the energy band $E_{+1}$ crosses the Fermi level only once, and there is simply one Fermi momentum and the accompanying Fermi velocity that contributes to the quartic coefficients. The situation changes drastically for $c >0$ where all three energy bands cross the Fermi level; we then find three different Fermi surfaces and the sum is over $i=-1_+,0,+1_+$ in Eqs.~\eqref{q1q2form}-\eqref{q3form}, with three different Fermi momenta $k_{{\rm F},0}$, $k_{{\rm F},-1_+}$, $k_{{\rm F},+1_+}$ and two different Fermi velocities $v_{{\rm F},0}$ and $v_{{\rm F},\pm 1_+}$. The leading order terms in the weak-coupling ($\mu/T_c \ll 1$) expressions for the quartic coefficients with $b=0$ are given as
\begin{eqnarray}
 q_1(T_c,c)= C_1 \begin{cases}
           \frac{v^2+2 c \mu}{c^2 \sqrt{v^2+4 c \mu}}  \frac{1}{T_c^2} &\text{ for } c<0 \\
            \frac{\mu^2}{v^3 T_c^2} &\text{ for } c=0\\
           \frac{v^2+2 c \mu+ 8 \sqrt{c \mu} \sqrt{v^2+4c \mu}}{c^2 \sqrt{v^2+4 c \mu}} \frac{1}{T_c^2} &\text{ for } c>0
          \end{cases}
\end{eqnarray}
$q_2(T_c,c)=q_1 (T_c,c)/2$, and
\begin{eqnarray}\label{q3oncaxis}
 q_3(T_c,c) =
 C_3 \begin{cases}
             \frac{1}{v^2 \sqrt{v^2+ 4 c \mu}} \log(\mu / T_c) & \text{ for }c<0\\
             \frac{1}{2 v^3} \log(\mu/T_c ) &\text{ for }c=0\\
             \frac{1}{3 v^2} \bigg(\frac{3}{\sqrt{v^2 +4 c \mu}}-\frac{2}{ \sqrt{c \mu}}  \bigg) \log(\mu/T_c ) &\text{ for }c>0 \:,
            \end{cases}
\end{eqnarray}
with $C_1= 1/ (2^{1/3} 3 ^{1/6} 15120 \pi)$, and $C_3= 3/(35 \pi^2) $ . The coefficients $q_1$ and $q_2$ are positive for all values of $c$ since all terms in the sum  $\sum_i {\eta}^{(1,2)}_i\frac{k_{{\rm F},i}^2}{v_{{\rm F},i}} $ arising from different Fermi surfaces are positive.  The situation is, however, different for the $q_3$ coefficient; for $c\leq 0$, there are  two Fermi surfaces that bring in two positive contributions to $q_3$ and yield a positive sign for $q_3$. On the other hand, when $c>0$, an additional Fermi surface appears and there is a competition between the contributions arising from different Fermi surfaces, i.e., a competition between $\tilde{\eta}_0/v_{{\rm F},0}$ and $\tilde{\eta}_{\pm 1_+}/v_{{\rm F},\pm 1_+}$. In other words, if we only consider contributions arising due to the expansions around $k_{{\rm F},\pm 1_+}$, which corresponds to $\tilde{\eta}_{{\pm 1}_+}/v_{{\rm F},{\pm1}_+}\propto3/ \sqrt{v^2+4 c\mu}$, the sign of $q_3$ would be positive. However, when we include the contribution from the expansion around $k_{{\rm F},0}$ which corresponds to $\tilde{\eta}_0/v_{{\rm F},0}\propto-2/ \sqrt{c\mu}$, it always dominates and ultimately leads to a negative sign for $q_3$. As a result, for $b=0$, the cyclic state is favored if $c\le 0$, and the ferromagnetic state is preferred when $c>0$.

As already mentioned earlier, for $\mu>0$ and $c<0$, we get positive contributions from two Fermi surfaces that originate from the same band $E_{+1}$. A similar situation also arises when one considers $\mu<0$ and $c>0$. In such a case, the $E_{-1}$ band intersects the Fermi level twice instead, forms two Fermi surfaces at two different momenta, and eventually leads to two positive contributions to $q_3$. One can easily check that the expressions for quartic coefficients stay invariant when one simultaneously transforms $\{\mu \rightarrow -\mu, b \rightarrow -b$, and $c \rightarrow -c\}$. This invariance suggests that the phase diagram for a positive chemical potential and the phase diagram for a negative chemical potential are related by two joint reflections, one around the $b$ axis and the other around the $c$ axis.

Finally, let us see how a finite parameter $b$ influences the superconducting state of the system. As shown in Fig.~\ref{fig:Fig3}, in the case of the positive chemical potential, the cyclic state is the preferred superconducting state for $c\le0$ and finite $b$. However, for positive $c$ we find that the superconducting state could be either the cyclic or the ferromagnetic state depending on the value of $b$. To understand the phase diagram qualitatively, let us assume again a finite, positive value of $c$ and $b=0$. As we already noted, for $b=0$ and $c>0$, the negative contribution arising due to the expansion around $k_{{\rm F},0}$ always dominates the positive $k_{{\rm F},\pm1_+}$ contribution and $q_3$ is consequently negative. However, coefficient $\tilde{\eta}_0/v_{{\rm F},0}$ gets progressively weaker with increasing $|b|$, and it even changes sign at a finite value of $b$. On the other hand, the two coefficients  $\tilde{\eta}_{\pm 1_+}/v_{{\rm F},\pm 1_+}$ for $b+c>0$ ($\tilde{\eta}_{+1_\pm}/v_{{\rm F},+1_\pm}$ for $b+c<0$), always stay positive.
As a consequence, one finds a phase boundary where the positive contribution from the Fermi surfaces at $k_{{\rm F},\pm 1_\pm }$ exceeds the negative contribution from near the Fermi momentum $k_{{\rm F},0}$, and any further increase of the absolute value of the parameter $b$ only yields the cyclic state as the superconducting ground state.

The curvature of the energy bands is thus found to be the ``knob" that can tune between the cyclic and the ferromagnetic superconducting state.

\begin{figure*}
 \includegraphics[width=2\columnwidth]{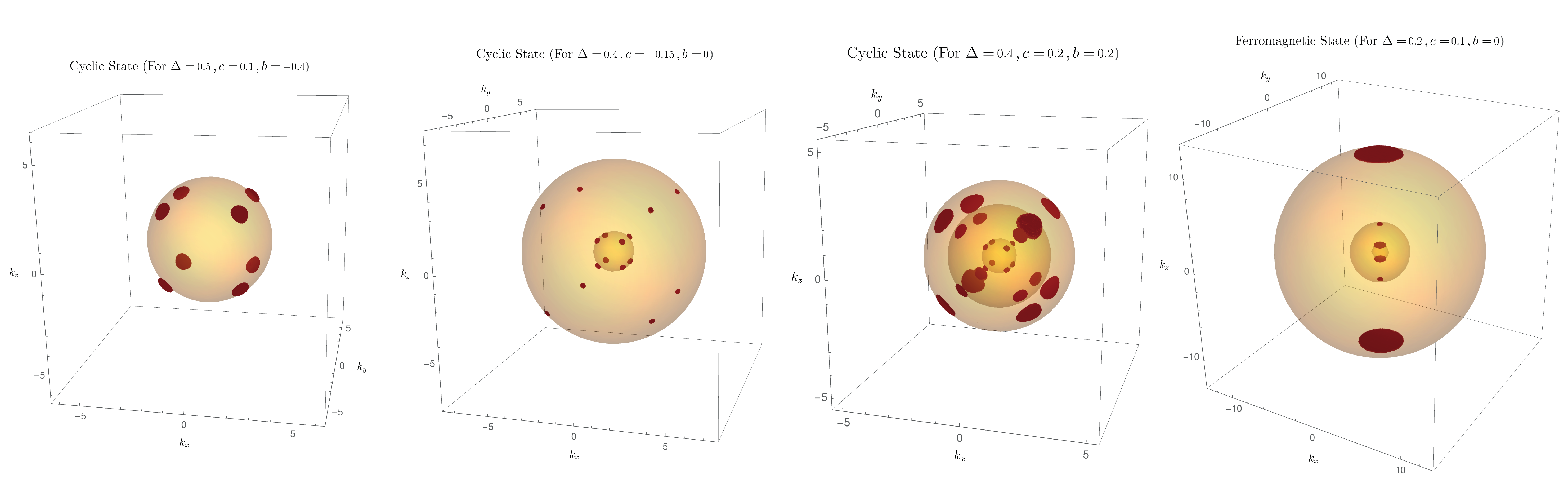}
\caption{BF surfaces that occur in the superconducting ground state of the fermionic system described by $H(\textbf{p})$ with $v=1$ and $\mu=1$.
Depending on the parameters $b$ and $c$, the cyclic state exhibits 8, 16, or 24 BF surfaces as shown in the first three panels from the left. However, the 
ferromagnetic state appears in a narrow region of the phase diagram where 6 BF surfaces emerge as shown in the fourth panel.
}
\label{fig:Fig4}
\end{figure*}

\subsection{Bogoliubov-Fermi surfaces}

In this section we analyze the energy spectrum of the Bogoliubov-de Gennes (BdG) quasiparticles which are described by the BdG-Hamiltonian $\sum_{\textbf{p}} \Psi^\dagger H_{\rm BdG} \Psi$ with
\begin{equation}
 H_{\rm BdG}=
 \begin{pmatrix}
  H(\textbf{p}) & \Delta_a \Gamma_a \\
  \Delta_a^\dagger \Gamma_a & - H(\textbf{p})
 \end{pmatrix}\:,
\end{equation}
where the Nambu spinor is given by ${\Psi(\omega_n,\textbf{p}) = \big(\psi(\omega_n,\textbf{p}), \mathcal{U} \psi^* (-\omega_n,-\textbf{p}) \big)}$.
We focus on the quasiparticle energy spectrum for the cyclic and the ferromagnetic state, and find that both of these TR-symmetry-breaking states exhibit extended regions in the momentum space where the energy vanishes. These regions are known as Bogoliubov-Fermi (BF) surfaces\cite{inflatednodes,agterberg2017bogoliubov,
bzduvsek2017robust,brydon2018bogoliubov,menke2019bogoliubov,lapp2019experimental,Setty2020, Link2,Herbut-LinkPRB2021} and can be determined by identifying the momentum points that satisfy $\det[H_{\rm BdG}(\mathbf{p})]=0$. The emergence of these BF surfaces in the present case is elaborated on in Appendix~\ref{ap:APP3}.\\

The mini BF surfaces that emerge for the cyclic and the ferromagnetic superconducting states are shown in Fig.~\ref{fig:Fig4}. Note that the number of BF surfaces is tied to the number of Fermi surfaces present in the normal state. If the superconducting ground state is the cyclic state, for example, then it displays 8, 16, or 24 BF surfaces depending on whether it has one, two, or three Fermi surfaces in the normal state. The BF surfaces in this case generally appear along the diagonals of a cube centered at the origin of the momentum space. This picture changes in the ferromagnetic state; this state emerges in a narrow region of the phase diagram where the normal state generally has three Fermi surfaces.  As a result, it typically exhibits six BF surfaces centered around the $z$-axis. We observe that despite of the triple-point fermion Hamiltonian's lack of inversion symmetry, the BF surfaces appear nevertheless, in accord with the general arguments presented in ref.~\onlinecite{Link2}.

\section{Conclusion}
\label{sec:discussion}

In conclusion, we have studied the $d$-wave superconductivity in the system of spin-1/2 fermions with triple-band crossings. In the weak-coupling limit and for a non-zero chemical potential, we have shown that the superconducting ground state prefers breaking of the time reversal symmetry, which leads to a competition between two distinct superconducting ground states that do so maximally: the cyclic and the ferromagnetic states. In contrast to previously studied examples, however, here we find that either of the two superconducting states can be the mean-field ground state, depending on the the parameters $b$ and $c$ in the single-particle Hamiltonian, which are directly related to the curvature of the energy bands. We have obtained the mean-field phase diagram below the superconducting critical temperature for finite chemical potential. In addition, the spectrum of the BdG quasiparticles in the cyclic and the ferromagnetic state is computed. Both of these states display multiple Bogoliubov-Fermi surfaces, in agreement with the general expectation for the multiband noncentrosymmetric superconductors that break time reversal\cite{Link2}.

Our main result is the surprising effect of the central $m=0$ band on the coefficients of the Ginzburg-Landau free energy. When this band is found to intersect the Fermi level, it yields a contribution to the key quartic term coefficient $q_3$ of the opposite sign of that from the other two bands, and that way it may overturn the overall sign of the $q_3$. Since the sign of this coefficient directly determines the time-reversal broken superconducting ground state, a negative $q_3$ leads to the appearance of the ferromagnetic state in the phase diagram. This state exhibits the maximal average magnetization, in addition to maximal breaking of time reversal dictated by the always-positive quartic term coefficient $q_2$. This is the first time, to the best of our knowledge, that such a possibility arises in a simple model that shows three-dimensional $d$-wave superconductivity.

\acknowledgements
JML is supported by the DFG grant No. LI 3628/1-1, and SM and IFH by the NSERC of Canada.

\appendix

\section{Tensor order and Cooper pairing}
\label{ap:cooperpairing}

The condensation of $\Delta_a=\langle \psi^{\rm T} \mathcal{U} \Gamma_a \psi \rangle$ indicates the onset of the $d$-wave superconductivity where $a=\{1,\cdots, 5\}$. These five components of ${\Delta}_a$ have the explicit form in terms of the original fermionic operators as
 \begin{eqnarray}
  \Delta_1 &=& 2 \rmi \langle a_{1,\downarrow} a_{1,\uparrow}+ a_{-1,\downarrow} a_{-1,\uparrow} \rangle \:, \\
  \Delta_2 &=& \frac{2 \rmi}{\sqrt{3}} \langle a_{-1,\downarrow} a_{1,\uparrow}+ a_{1,\downarrow} a_{-1,\uparrow} + 2 a_{0,\downarrow} a_{0,\uparrow} \rangle  \:, \\
  \Delta_3 &=& \frac{2 \rmi}{\sqrt{3}} \langle - a_{0,\downarrow} a_{1,\uparrow} -  a_{1,\downarrow} a_{0,\uparrow} +  a_{-1,\downarrow} a_{0,\uparrow} +  a_{0,\downarrow} a_{-1,\uparrow} \rangle \:, \\
  \Delta_4 &=& \frac{2}{\sqrt{2}} \langle a_{0,\downarrow} a_{1,\uparrow} + a_{1,\downarrow} a_{0,\uparrow} + a_{-1,\downarrow} a_{0,\uparrow} + a_{0,\downarrow} a_{-1,\uparrow} \rangle \:, \\
  \Delta_5 &=& 2 \langle -a_{1,\downarrow} a_{1,\uparrow} + a_{-1,\downarrow} a_{-1,\uparrow} \rangle
  \:,
 \end{eqnarray}
with the coefficients which may be recognized as related to the usual Clebsch-Gordan coefficients. In the case of a reduced cubic symmetry,
$(\Delta_1,\Delta_2)$ would belong to the E-representation, while $(\Delta_3,\Delta_4,\Delta_5)$ would belong to the $T_{2g}$-representation.

In contrast, the standard $s$-wave order parameter corresponds to
 \begin{equation}
  \Delta_s=\langle \psi^{\rm T} \mathcal{U} \psi \rangle
  =2 \rmi \langle a_{1,\downarrow} a_{-1,\uparrow} + a_{-1,\downarrow} a_{1,\uparrow} - a_{0,\downarrow} a_{0,\uparrow} \rangle
  \:.
 \end{equation}

\section{Potential Ground States}
\label{ap:APPB}
Since the irreducible representations of the $SO(3)$ of given dimension are unique, the tensor $d$-wave order parameter is entirely equivalent to a quantum state in the spin-$2$ Hilbert space. The five real Gell-Mann matrices in Eq.~\eqref{eq5b} which form the basis in the irreducible tensor (symmetric traceless matrix) space, under the $SO(3)$ rotations transform exactly as the following states, given as linear combinations in  the standard basis:
\begin{align}
|M_1 \rangle &= \dfrac{1}{\sqrt{2}}\Big( |-2 \rangle + |2 \rangle \Big)   \\
|M_2 \rangle &= |0 \rangle  \\
|M_3 \rangle &= \dfrac{1}{\sqrt{2}}\Big( |-1 \rangle - |1 \rangle \Big)  \\
|M_4 \rangle &= \dfrac{i}{\sqrt{2}}\Big( |-1 \rangle + |1 \rangle \Big)  \\
|M_5 \rangle &= \dfrac{i}{\sqrt{2}}\Big( |-2 \rangle - |2 \rangle \Big)
\end{align}
where, $J_z|m\rangle=m|m\rangle$ and $m=0,\pm1,\pm2$. For the five-component $d$-wave superconducting order parameter, the free energy to the quartic order can be shown to have the following form \cite{mermin1974d,PhysRevLett.120.057002}
\begin{equation}
\begin{split}
F(\Delta) &= r \Delta^* _a \Delta_a+  q_1 (\Delta^* _a \Delta_a)^2 +q_2 |\Delta_a \Delta_a|^2 \\ &+ \frac{q_3}{2} \Tr \big(\phi^\dagger \phi \phi^\dagger \phi \big) + \mathcal{O}(\Delta^6)
\end{split}
\end{equation}

Mermin was the first to directly minimize the above free energy for the  $d$-wave order parameter \cite{mermin1974d}. In a related approach, in the context of the spinor Bose-Einstein condensates Kawaguchi and Ueda employed  an elegant and general symmetry-based approach based on Michel's theorem. \cite{KawaguchiUeda}
They found that among all the  ground state candidates with some residual symmetry, the real states, the ferromagnetic state, and the cyclic state are the only true minima of the Ginzburg-Landau free energy, each one winning in a different part of the parameter space. In the above ``real" (TR invariant)  basis $|M_i\rangle$ these states are given by:
\begin{align}
{\vec{\Delta}}_{\rm real} &= \big({\Delta}_1,{\Delta}_2,0,0,0\big) \\
{\vec{\Delta}}_{\rm ferro} &= \dfrac{\Delta}{\sqrt{2}}\big(1,0,0,0,i \big) \\
{\vec{\Delta}}_{\rm cyclic} &= \dfrac{\Delta}{\sqrt{2}}\big(1,i,0,0,0 \big)
\end{align}
Note that ${\vec{\Delta}}_{\rm ferro}$ is simply proportional to the state $ |2 \rangle$, and therefore has the maximal magnetization. ${\vec{\Delta}}_{\rm cyclic}$, on the other hand, has the average magnetization of zero, as may be easily checked. Both states break the TR maximally, i.e. $\Delta_i \Delta_i =0$.
In terms of three dimensional traceless matrices these states are equivalent to the following matrix order parameters:
\begin{align}
{\phi}_{\rm real}= {\Delta}_1M_1+{\Delta}_2M_2\\
{\phi}_{\rm ferro}= \dfrac{\Delta}{\sqrt{2}}\big(M_1+iM_5 \big)\\
{\phi}_{\rm cyclic}= \dfrac{\Delta}{\sqrt{2}} \big( M_1+iM_2\big)
\end{align}

\section{Computation of the coefficients}
\label{sec:computation}
Here we outline the  computation of the coefficients $q_{1,2,3}$ in the Ginzburg-Landau free energy. As usual, one first obtains the mean-field expression for the free energy by integrating out the fermionic degrees of freedom for a constant order parameter, and then expands the mean-field free energy to the quartic order in the order parameter. These steps lead to

\begin{align}
 \label{strong1} F_2(\Delta) &= \big(\frac{1}{g}\delta_{ab} -\frac{1}{2}K_{ab} \big) \Delta^*_a\Delta_b,\\
 \label{strong2} F_4(\Delta) &= \frac{1}{4}K_{abcd} \Delta^*_a\Delta_b\Delta^*_c\Delta_d
\end{align}
with
\begin{align}
 \label{strong3} K_{ab} = \mbox{tr} \int_Q^\Lambda {}&G_0(\omega_n,\mu, \textbf{q})\Gamma_a G_0(-\omega_n,\mu,\textbf{q})\Gamma_b \:,\\
 \nonumber K_{abcd}= \mbox{tr}\int_Q^\Lambda {}&G_0(\omega_n,\mu,\textbf{q}) \Gamma_a G_0(-\omega_n,\mu,\textbf{q})\Gamma_b \\
 \label{strong4} &\times G_0(\omega_n,\mu,\textbf{q}) \Gamma_c G_0(-\omega_n,\mu,\textbf{q})\Gamma_d
 \:.
\end{align}
The measure of the integrals are given by
\begin{equation}
 \int_Q^\Lambda :=T \sum_{n \in \mathbb{Z}} \int_{\textbf{q}}^\Lambda := T \sum_{n \in \mathbb{Z}} \int_{q\leq \Lambda} \frac{d^3 q}{(2 \pi)^3}
 \:,
\end{equation}
with the fermionic Matsubara frequency $\omega =(2n+1)\pi T$, the temperature $T$, and the ultraviolet cutoff $\Lambda \gg \mu,T$. The fermionic Green's function is defined as
\begin{equation}
 G_0(\omega_n,\mu,\textbf{p})
 =
 \big(\rmi \omega_n -H(\textbf{p}) \big)^{-1}
 \:.
\end{equation}
To determine the coefficients $q_{1,2,3}$ we insert the states $\Delta_1=\Delta(0,1,0,0,0)$, $\Delta_2=\frac{\Delta}{\sqrt{2}}(1,\rmi,0,0,0)$, and $\Delta_3=\frac{\Delta}{\sqrt{2}}(0,0,1,\rmi,0)$ in Eq.~\eqref{strong4} and thus obtain the following matching conditions:
\begin{eqnarray}
 F_4(\Delta_1)&=&(q_1+q_2+q_3) \Delta^4 \\
 F_4(\Delta_2)&=& \big(q_1 + \frac{2}{3} q_3 \big) \Delta^4\\
 F_4(\Delta_3)&=& (q_1+q_3)\Delta^4
 \:.
\end{eqnarray}
\begin{widetext}
 \subsection{Explicit expressions for the coefficients}
 \label{ap:APP2}
Here, we have presented the expressions for the coefficients with the setting of $v=1$. In such scenario, the quadratic coefficient $r$ is given by
 \begin{equation}
r(g,\mu,T,\Lambda) = \dfrac{1}{g} -\dfrac{1}{2}K_{11}(T,\mu,\Lambda)
\end{equation}
 where,
 \begin{equation}
 \begin{split}
 K_{11} &= \int_{Q}^{\Lambda} \Big(4 (p^8 (b+c)^2 (2 b^2+10 b c+15 c^2)-p^6 (2 \mu  (b+c) (7 b^2+30 b c+30 c^2)+4 b^2+14 b c+15 c^2)+p^4 (\omega ^2 (17 b^2\\
 &+40 b c+30 c^2)+\mu  (37 b^2 \mu +2 b (60 c \mu +7)+30 c (3 c \mu +1))+2)+5 p^2 (\omega ^2 (-8 b \mu -12 c \mu +1)-\mu ^2 (8 b \mu +12 c \mu +3))\\
 &+15 (\mu ^2+\omega ^2)^2)\Big)/\Big(15 (c p^2-\mu -i \omega ) (c p^2-\mu +i \omega ) (p^2 (b+c)-\mu +p-i \omega ) (p (p (b+c)-1)-\mu -i \omega )\\
 &\times (p^2 (b+c)-\mu +p+i \omega ) (p (p (b+c)-1)-\mu +i \omega )\Big)
  \:.
 \end{split}
\end{equation}
The transition temperature for different $b$ and $c$ parameters can be obtained from $r(g,\mu,T_c,\Lambda)=0$. Here we refrain from going into details of the result for the transition temperature, which ultimately has the standard weak-coupling form, and focus on the nature of the superconducting ground state.
This requires the computation of the quartic terms in the Ginzburg-Landau expansion. For a finite chemical potential $\mu$, the coefficients of the three independent quartic terms are given by
 \begin{equation}
\begin{split}
&q_1(T,b,c)= \int_{Q}^{\Lambda} \Big(2 p^2 (b^2 (b+c)^4 (8 b^2+48 c b+63 c^2) p^{14}+(b+c)^2 (-32 b^4-176 c b^3-201 c^2 b^2-2 (b+c) (b+3 c) (40 b\\
&+63 c) \mu  b^2+42 c^3 b+105 c^4) p^{12}+(3 (101 \mu ^2 b^6+4 \mu  (143 c \mu +20) b^5+2 (c \mu  (578 c \mu +195)+8) b^4+8 c (c \mu  (125 c \mu +67)\\
&+10) b^3+c^2 (c \mu  (315 c \mu +16)+87) b^2-12 c^3 (35 c \mu +1) b-42 c^4 (5 c \mu +1))-b^2 (b+c)^2 (47 b^2+30 c b-63 c^2) \omega ^2) p^{10}\\&+(-572 \mu ^3 b^5-\mu ^2 (2312 c \mu +585) b^4-24 \mu  (c \mu  (125 c \mu +67)+10) b^3-2 (9 c \mu  (2 c \mu  (35 c \mu +2)+29)+16) b^2\\
&+4 c (9 c \mu  (70 c \mu +3)-20) b+(124 \mu  b^5+(88 c \mu +169) b^4+8 c (107-36 c \mu ) b^3-252 c^2 (c \mu -6) b^2+1008 c^3 b+105 c^4) \omega ^2\\
&+21 c^2 (3 c \mu  (25 c \mu +8)+1)) p^8+(-b^2 (118 b^2+204 c b+63 c^2) \omega ^4-(44 \mu ^2 b^4+8 \mu  (107-36 c \mu ) b^3+(197-378 c \mu  (c \mu \\
&-8)) b^2+12 c (252 c \mu +61) b+84 c^2 (5 c \mu +7)) \omega ^2+\mu  (578 \mu ^3 b^4+4 \mu ^2 (375 c \mu +134) b^3+3 \mu  (c \mu  (315 c \mu +16)+87) b^2\\
&-4 (9 c \mu  (70 c \mu +3)-20) b-42 c (2 c \mu  (25 c \mu +9)+1))+8) p^6+3 ((68 \mu  b^3+2 (21 c \mu +86) b^2+252 c b-35 c^2) \omega ^4\\
&+(2 \mu  (-16 \mu ^2 b^3-42 \mu  (c \mu -6) b^2+2 (252 c \mu +61) b+7 c (15 c \mu +28))+25) \omega ^2+\mu ^2 (\mu  (-100 \mu ^2 b^3-2 \mu  (63 c \mu +2) b^2\\
&+12 (35 c \mu +1) b+21 c (25 c \mu +8))+7)) p^4+21 (-3 b^2 \omega ^6+(10 c \mu -3 b (b \mu +12) \mu +2) \omega ^4+\mu ^2 (-20 c \mu +3 b (b \mu -16) \mu \\
&-28) \omega ^2+3 \mu ^4 (-10 c \mu +b (b \mu -4) \mu -2)) p^2+105 (\mu -\omega ) (\mu +\omega ) (\mu ^2+\omega ^2)^2)\Big)/\Big(315 ((b+c) p^2+p-\mu -i \omega )^2\\
&\times (c p^2-\mu +i \omega )^2 ((b+c) p^2+p-\mu +i \omega )^2 (p ((b+c) p-1)-\mu +i \omega )^2 (-c p^2+\mu +i \omega )^2 (-(b+c) p^2+p+\mu +i \omega )^2\Big) ,
\end{split}
\end{equation}
 \begin{equation}
\begin{split}
&q_2(T,b,c)= \int_{Q}^{\Lambda} \Big(p^2 (b^2 (b+c)^4 (8 b^2+24 c b+21 c^2) p^{14}+(b+c)^2 (-56 \mu  b^5-2 (109 c \mu +16) b^4-8 c (36 c \mu +13) b^3\\
&-3 c^2 (42 c \mu +65) b^2-210 c^3 b-105 c^4) p^{12}+(3 (55 \mu ^2 b^6+4 \mu  (65 c \mu +14) b^5+2 (5 c \mu  (46 c \mu +29)+8) b^4+8 c (c \mu  (45 c \mu \\
&+88)+7) b^3+5 c^2 (3 c \mu  (7 c \mu +64)+25) b^2+4 c^3 (175 c \mu +37) b+70 c^4 (3 c \mu +1))-b^2 (b+c)^2 (5 b^2-6 c b-21 c^2) \omega ^2) p^{10}\\
&+(4 \mu  (\omega ^2-65 \mu ^2) b^5-(5 (184 c \mu +87) \mu ^2+(56 c \mu -19) \omega ^2) b^4+8 (-2 c (9 c \mu +11) \omega ^2-3 \mu  (c \mu  (45 c \mu +88)+7)) b^3\\
&+2 (3 c (-14 c (c \mu +8) \omega ^2-5 \mu  (2 c \mu  (7 c \mu +72)+25))-16) b^2-4 c (3 c (56 c \omega ^2+\mu  (350 c \mu +111))+14) b\\
&-105 c^2 (c (c \omega ^2+\mu  (15 c \mu +8))+1)) p^8+(-b^2 (34 b^2+60 c b+21 c^2) \omega ^4+(28 \mu ^2 b^4+16 \mu  (9 c \mu +11) b^3+(42 c \mu  (3 c \mu +32)\\
&-23) b^2+36 c (56 c \mu +5) b+84 c^2 (5 c \mu +1)) \omega ^2+\mu  (230 \mu ^3 b^4+4 \mu ^2 (135 c \mu +176) b^3+15 \mu  (3 c \mu  (7 c \mu +64)+25) b^2\\
&+4 (3 c \mu  (350 c \mu +111)+14) b+210 c (2 c \mu  (5 c \mu +3)+1))+8) p^6+3 ((20 \mu  b^3+2 (7 c \mu -48) b^2-84 c b+35 c^2) \omega ^4\\
&-(2 \mu  (8 \mu ^2 b^3+14 \mu  (c \mu +8) b^2+6 (56 c \mu +5) b+7 c (15 c \mu +4))-3) \omega ^2+\mu ^2 (\mu  (-36 \mu ^2 b^3-6 \mu  (7 c \mu +40) b^2-4 (175 c \mu \\
&+37) b-35 c (15 c \mu +8))-35)) p^4+21 (-b^2 \omega ^6+(2-\mu  (10 c+b (b \mu -12))) \omega ^4+\mu ^2 (20 c \mu +b (b \mu +32) \mu +4) \omega ^2\\
&+\mu ^4 (30 c \mu +b (b \mu +20) \mu +10)) p^2-105 (\mu -\omega ) (\mu +\omega ) (\mu ^2+\omega ^2)^2)\Big)/\Big(315 ((b+c) p^2+p-\mu -i \omega )^2 \\
&\times(c p^2-\mu +i \omega )^2 ((b+c) p^2+p-\mu +i \omega )^2 (p ((b+c) p-1)-\mu +i \omega )^2 (-c p^2+\mu +i \omega )^2 (-(b+c) p^2+p+\mu +i \omega )^2\Big) ,
\end{split}
\end{equation}
 \begin{equation}
\begin{split}
& q_3(T,b,c)= \int_{Q}^{\Lambda} \Big(105 c^8 p^{16}+8 b^7 (c p^2-\mu ) p^{14}-840 c^7 \mu  p^{14}+b^6 (95 (\mu -c p^2)^2+49 \omega ^2) p^{12}+35 c^6 (12 (7 \mu ^2+\omega ^2)-7 p^2) p^{12}\\
&+8 b^5 (c p^2-\mu ) (55 c^2 p^4-(110 c \mu +3) p^2+55 \mu ^2+44 \omega ^2) p^{10}+210 c^5 \mu  (7 p^2-4 (7 \mu ^2+3 \omega ^2)) p^{10}+b^4 (203 \omega ^4+(1222 c^2 p^4 \\
&-(2444 c \mu +111) p^2+1222 \mu ^2) \omega ^2+25 (\mu -c p^2)^2 (43 c^2 p^4-(86 c \mu +9) p^2+43 \mu ^2)) p^8+35 c^4 (5 p^4-15 (7 \mu ^2+\omega ^2) p^2\\
&+6 (35 \mu ^4+30 \omega ^2 \mu ^2+3 \omega ^4)) p^8-140 c^3 \mu  (5 p^4-5 (7 \mu ^2+3 \omega ^2) p^2+6 (\mu ^2+\omega ^2) (7 \mu ^2+3 \omega ^2)) p^6+8 b^3 (c p^2-\mu ) (190 c^4 p^8 \\
&-c^2 (760 c \mu +91) p^6+(c (303 c \omega ^2+2 \mu  (570 c \mu +91))+3) p^4-((760 c \mu +91) \mu ^2+3 (202 c \mu +23) \omega ^2) p^2\\
&+(\mu ^2+\omega ^2) (190 \mu ^2+113 \omega ^2)) p^6+b^2 (259 \omega ^6+(1771 c^2 p^4-2 (1771 c \mu +41) p^2+1771 \mu ^2) \omega ^4+(2765 c^4 p^8-28 c^2 (395 c \mu \\
&+39) p^6+3 (14 c \mu  (395 c \mu +52)+25) p^4-28 \mu ^2 (395 c \mu +39) p^2+2765 \mu ^4) \omega ^2+(\mu -c p^2)^2 (1253 c^4 p^8-2 c^2 (2506 c \mu +561) p^6\\
&+3 (2 c \mu  (1253 c \mu +374)+55) p^4-2 \mu ^2 (2506 c \mu +561) p^2+1253 \mu ^4)) p^4-7 c^2 (5 p^6-6 (25 \mu ^2+3 \omega ^2) p^4+15 (35 \mu ^4+30 \omega ^2 \mu ^2\\
&+3 \omega ^4) p^2-60 (\mu ^2+\omega ^2)^2 (7 \mu ^2+\omega ^2)) p^4+14 c \mu  (5 p^6-2 (25 \mu ^2+9 \omega ^2) p^4+15 (\mu ^2+\omega ^2) (7 \mu ^2+3 \omega ^2) p^2-60 (\mu ^2+\omega ^2)^3) p^2\\
&+8 b (c p^2-\mu ) (70 c^6 p^{12}-105 c^4 (4 c \mu +1) p^{10}+6 c^2 (35 c (c \omega ^2+\mu  (5 c \mu +2))+6) p^8-(c (7 c (120 c \mu +19) \omega ^2\\
&+2 \mu  (35 c \mu  (20 c \mu +9)+36))+1) p^6+(210 c^2 \omega ^4+(14 c \mu  (90 c \mu +19)+25) \omega ^2+6 \mu ^2 (35 c \mu  (5 c \mu +2)+6)) p^4-7 (\mu ^2+\omega ^2)\\
&\times(4 \omega ^2+15 \mu  (4 c \mu ^2+\mu +4 c \omega ^2)) p^2+70 (\mu ^2+\omega ^2)^3) p^2+105 \omega ^8-35 (p^2-12 \mu ^2) \omega ^6+7 (p^4-45 \mu ^2 p^2+90 \mu ^4) \omega ^4\\
&-35 \mu ^2 (p^2-3 \mu ^2) (p^2-\mu ^2)^2+(-13 p^6+126 \mu ^2 p^4-525 \mu ^4 p^2+420 \mu ^6) \omega ^2\Big)/\Big(105 ((b+c) p^2+p-\mu -i \omega )^2 (c p^2-\mu +i \omega )^2 \\
&\times((b+c) p^2+p-\mu +i \omega )^2 (p ((b+c) p-1)-\mu +i \omega )^2 (-c p^2+\mu +i \omega )^2 (-(b+c) p^2+p+\mu +i \omega )^2\Big) ,
\end{split}
\end{equation}
At finite temperatures, all of these integrals are finite upon introducing an UV-cutoff $\Lambda$ and can be evaluated by performing the Matsubara sums and momentum integrations numerically. Generally, the most significant contributions at low temperature come from the regions near the normal Fermi surfaces. An approximate ansatz for integrands of $q_{1,2,3}$ can be constructed from first calculating the zero-temperature and the finite-temperature Matsubara sums, then expanding them around the Fermi surfaces. Whereas the first one captures the divergent behavior of integrand at zero temperature,  the latter can be used to extract the temperature dependence. Our analysis shows that the coefficients $q_{1,2}$ always stay positive, and hence sign of the coefficient $q_3$ selects the superconducting ground state.

\end{widetext}

 \section{Emergence of the BF surface}
 \label{ap:APP3}
In this appendix we briefly explain why BF surfaces emerge in the TR-breaking superconducting states.
To this end, we use the iterative procedure introduced in ref.~\onlinecite{Link2}, where an effective Hamiltonian was found for the energy band crossing the chemical potential  at a fixed momentum.

In the present case, the $12\times 12$ BdG Hamiltonian
with
\begin{equation}
 H_{\rm BdG}=\begin{pmatrix}
              \mathbb{1}_{2\times2} \otimes (H_0(\textbf{p})-\mu) & \Delta_a (\mathbb{1}_{2\times 2} \otimes \gamma_a)\\
              \Delta_a^\dagger (\mathbb{1}_{2\times 2} \otimes \gamma_a) & -\mathbb{1}_{2\times2} \otimes (H_0(\textbf{p})-\mu)
             \end{pmatrix}
\end{equation}
can be reduced to a $6\times6$ Hamiltonian upon rearranging the blocks of the matrix to $H_{\rm BdG}=\mathbb{1}_{2\times2} \otimes h_{\rm BdG}$. $h_{\rm BdG}$ is defined as
\begin{equation}
 h_{\rm BdG}=\begin{pmatrix}
              H_{0}(\textbf{p}) -\mu & \gamma \\
              \gamma^\dagger & - [H_{0}(\textbf{p}) -\mu]
             \end{pmatrix}
             \:,
\end{equation}
where the pairing matrix is given by $\gamma = \Delta_a \gamma_a$. The eigenstates of the Hamiltonian $H_0(\textbf{p})$ are defined as $\phi_{\pm 1}(\textbf{p})$ and $\phi_0(\textbf{p})$. We further use the following properties of the pairing matrices and the eigenstates, namely:
\begin{equation}
 U_0^\dagger \gamma_a U_0 = \gamma_a^{\rm T} \:,
\end{equation}
and
\begin{eqnarray}
 \phi_{-1}(\textbf{p})&=& U_0 \phi_{+1}(\textbf{p})^* \\
  \phi_{1}(\textbf{p})&=& U_0 \phi_{-1}(\textbf{p})^* \\
   \phi_{0}(\textbf{p})&=& U_0 \phi_{0}(\textbf{p})^*
   \:.
\end{eqnarray}
Hence, the BdG-Hamiltonian for $H(\textbf{p})$ which describes fermions with an angular momentum of $L=1$ and a spin of $S=1/2$ can be cast into the following form:
 \begin{eqnarray}
 h_{\rm BdG}(\textbf{p})
 =\begin{pmatrix}
   E_1 & X & 0 & A & 0 & B\\
   \bar{X} & - E_1 & \bar{D} & 0 & \bar{C} & 0 \\
   0 & D & E_0 & Y & 0 & A \\
   \bar{A} & 0 & \bar{Y} & - E_0 & \bar{D} & 0 \\
   0 & C & 0 & D & E_{-1} & X \\
   \bar{B} & 0 & \bar{A} & 0 & \bar{X} & - E_{-1}
  \end{pmatrix}
\:,
 \end{eqnarray}
 where the intra- and interband couplings are defined as
 \begin{eqnarray}
  X &=& \phi_1^\dagger(\textbf{p}) \gamma \phi_1(\textbf{p}) \\
  Y &=& \phi_0^\dagger(\textbf{p}) \gamma \phi_0(\textbf{p}) \\
  A &=& \phi_1^\dagger(\textbf{p}) \gamma \phi_0(\textbf{p}) \\
  B &=& \phi_1^\dagger(\textbf{p}) \gamma \phi_{-1}(\textbf{p}) \\
  C &=& \phi_{-1}^\dagger(\textbf{p}) \gamma \phi_{+1}(\textbf{p})\\
  D &=& \phi_0^\dagger(\textbf{p}) \gamma \phi_{+1}(\textbf{p}) \:.
 \end{eqnarray}
In the normal state, there can be up to three Fermi surfaces, since the curvature of the energy bands determines which energy bands intersect the Fermi level. Here, let us assume that the energy band $E_{+1}$ intersects the Fermi level once and study whether a BF surface emerges in the superconducting state. After integrating out the energy bands that are far above the chemical potential, the energy band closest to the chemical potential is described by the following effective Hamiltonian in 2nd order perturbation theory:
\begin{equation}
 H_{ef}=\begin{pmatrix}
         E_1 & X \\
         \bar{X} & -E_1
         \end{pmatrix}
         -
         \begin{pmatrix}
         -\frac{|A|^2}{E_0} -\frac{|B|^2}{E_1} & 0 \\
         0 & \frac{|D|^2}{E_0} + \frac{|C|^2}{E_1}
        \end{pmatrix}
\end{equation}
The interband pairing between the different states introduces both a shift in the momentum and in the energy of the energybands of the BdG quasiparticles when $A\neq D$ and $C\neq B$, which is the case if time-reversal symmetry is broken. The shift in the energy leads to the emergence of the BF surfaces.

%\bibliography{ref_chiral}


\end{document}